\documentstyle[12pt,psfig,aps]{revtex}

\newcounter{figcounter}
\def\T{E_{kin}}

\def\<{\langle}
\def\>{\rangle}
\def\({\left(}
\def\){\right)}
\def\MeV{\nobreak\,\mbox{MeV}}
\def\GeV{\nobreak\,\mbox{GeV}}
\def\fm{\nobreak\,\mbox{fm}}
\def\mb{\nobreak\,\mbox{mb}}
\def\L{{\cal L}}
\def\M{{\cal M}}
\def\epm{e^+e^-}
\newcommand{\be}{\begin{eqnarray}}
\newcommand{\ee}{\end{eqnarray}}
\newcommand{\bes}{\begin{eqnarray*}}
\newcommand{\ees}{\end{eqnarray*}}
\begin{document}
\title{Intermediate mass excess of dilepton production in
heavy ion collisions at \mbox{BEVALAC} energies\footnote{Supported by
Graduiertenkolleg Theoretische und Experimentelle
Schwerionenphysik, GSI, BMBF and DFG}}
\author{C.~Ernst\footnote{e-mail:
ernst@th.physik.uni-frankfurt.de},
S.~A.~Bass, M.~Belkacem,\\
H.~St\"{o}cker and W.~Greiner\\
{\normalsize \it Institut f\"{u}r Theoretische Physik, 
J. W. Goethe-Universit\"{a}t}\\
{\normalsize \it D-60054 Frankfurt am Main, Germany}}
\date{\today}
\maketitle
\begin{abstract}
Dielectron mass spectra are examined for
various nuclear
reactions recently measured by the DLS
collaboration.
A detailed description is given of all dilepton
channels 
included in the transport model UrQMD 1.0, 
i.e. Dalitz decays of
$\pi^0,\eta,\omega,\eta'$ mesons and of the
$\Delta(1232)$ resonance, direct decays of vector
mesons and $pn$ bremsstrahlung. 
The microscopic calculations reproduce data for
light systems fairly well, but tend to
underestimate the data in $pp$ at high energies
and in $pd$ at low energies.
These conventional sources, however,
cannot explain the recently reported enhancement for
nucleus-nucleus collisions in the mass region 
$0.15\GeV\le M_{\epm}\le 0.6\GeV$.
Chiral scaling and $\omega$ meson broadening in
the medium are investigated as a source of
this mass excess. They also cannot explain the recent DLS data. 
\end{abstract}
\newpage
\section{Introduction}
\label{secti}
Dileptons have been proposed in the late 70's as
penetrating probes \cite{feinberg76a}
of hot and dense nuclear systems.
They are presumably created in
all stages of heavy ion reactions by several distinct
mechanisms. Once produced, they
practically do not interact with the surrounding hadronic
matter. Low mass dileptons are of particular
interest (see e.g. 
\cite{gale87b,wolf90a,titov95a,cassing95a,ko96a,bratko96a}).
They can reveal information of the
$hadronic$ properties in the reaction zone.
Several experiments have focused on
low mass lepton pairs:
the DLS spectrometer at the BEVALAC
\cite{porter97a,wilson97a}, the CERES and HELIOS detectors
at the SPS at CERN \cite{drees96a}. The 
dilepton spectrometers HADES at SIS at GSI \cite{stroth95a} and PHENIX at
RHIC in BNL \cite{gregory94a} are under construction.
The most striking result of the high energy dilepton
programs so far is the observed enhancement
in heavy systems at low invariant masses  
as compared to "conventional" hadronic cocktails and 
models.
A dropping mass \cite{brown91a,hatsuda95a} or
dissolving  spectral function
\cite{rapp97a,cassing97a} of
the $\rho$ meson have been offered in attempt
to explain these data.
Recently, a systematic measurement of dilepton
production at BEVALAC energies has been
published. Data are available for elementary $pp$
and $pd$ collisions as well as for
nucleus-nucleus collisions \cite{porter97a,wilson97a}.

The aim of the present work is to investigate dilepton
production within the microscopic $n$-body transport model 
UrQMD.
The outline of the paper is as follows: In
section \ref{sectii} 
a brief survey of the UrQMD model is given. A more
detailed description of the implemented dilepton
production mechanisms follows.  Section \ref{sectiii}
contains the calculations of the elementary $pp$ and $pd$
dilepton cross sections in comparison to recent
DLS measurements. Section \ref{sectiv} shows
mass spectra for heavier systems. Summary and concluding remarks 
are given in section
\ref{sectv}.
\section{Production of resonances and dileptons}
\label{sectii}
\subsection{Hadron production in the UrQMD model}
\label{sectii-i}
The UrQMD model is based on the quantum molecular dynamics
concept
\cite{aichelin86a,bass95c,winckelmann96a}.
The Hamilton's equations of motion 
are solved for gaussian wave packets. The
model allows for the production of all
established meson and baryon resonances up to
about $2\GeV$ with all
corresponding isospin projections  and antiparticle-states.
The collision term
describes particle production by 
resonant excitation channels
and, for higher masses, within a string fragmentation
scheme. The UrQMD model is designed to cover consistently
the whole range of bombarding energies per nucleon from $200\MeV$ to
$200\GeV$. For dilepton production at \mbox{BEVALAC}
energies, the resonant production of neutral
mesons is most important. A detailed description of the model can
be found in refs.
\cite{bass2appear,bleicher2appear}.

The formation of light mesons
at low energies is modelled as a
multistep process that proceeds via intermediate
heavy baryon and meson resonances and their subsequent decay
\cite{winckelmann95a}.
The resonance parameters (pole masses, widths
and branching ratios) are taken from
\cite{PDG96}, but large
uncertainties of these parameters are used to obtain a consistent fit
to cross section data. For example, the production
of $\omega$ mesons is described in the UrQMD model by the formation and
the decay of the $N^\star(1900)$ resonance.
 It decays to 35\% into $N\pi$
and to 55\% into $N\omega$. As suggested in
ref. \cite{batinic95a}, the $\eta$
production proceeds not only via
$N^\star(1535)$, but invokes also nucleon resonances
with masses
1650, 1700, 1710 and 2080\MeV.
A full list of the UrQMD resonance parameters is
published in ref. \cite{bass2appear}.

Broad resonances have mass dependent decay
widths.
For the resonance cross sections in baryon-baryon,
meson-baryon and meson-meson collisions 
conventional Breit-Wigner parametrisations are used
with mass dependent widths.
At higher energies, the resonant particle production
does no longer predict the observed cross
sections. There, the string picture, described in
\cite{bleicher2appear}, is employed.

Fig. \ref{ppc} shows calculations of 
exclusive ($pp\to m pp$) and inclusive ($pp\to
mX$) cross sections for the production of
neutral mesons $m=\pi^0,\eta, \rho^0, \omega$. Recent
data on the exclusive $\eta$ production just
above threshold
are reproduced reasonably well \cite{calen96a}.
Note that above 3.5~GeV the exclusive cross sections become
less important, because the string channels open and allow
for multiple resonance production.
For the present
article the relevant $\sqrt{s}$ values are below
4~GeV, where so far no data on the inclusive channels
are available.
\subsection{Dilepton radiation in UrQMD}
\label{sectii-ii}
Dileptons can be produced in hadronic decays and
collisions. The mechanisms that
are expected to dominate in the low mass
region (with invariant dilepton masses below
1~GeV) are the Dalitz decays ($A\to B\epm$)
of neutral mesons $\pi^0,\eta,\eta',\omega$, as well as the
Dalitz decay of the 
$\Delta(1232)$ resonance. Around their
pole masses the direct decays ($A\to \epm$) of the vector
mesons $\rho^0,\omega$ and $\phi$ are expected to dominate the
spectrum. These modes are of special interest:
the invariant mass of the dilepton equals
the vector meson mass in the hot and dense medium.
Also included in the present model calculation is the  $pn$
bremsstrahlung.
It contributes especially at low bombarding energies.
In accord with ref. \cite{wolf90a}, other
sources like Dalitz decays of heavier resonances,
$pp$, $\pi N$ bremsstrahlung etc. are neglected
for the reactions considered here. 
Direct dilepton production in binary collisions, e.g.
$\pi^+\pi^-,K^+K^-\to \epm$ or $\pi\rho\to
\epm$, is not evaluated. In the framework of the present model
this would correspond to a partial double counting: these
reactions are supposedly included as explicit multi-step
processes (e.g. $\pi\rho\to\phi\to \epm$).
This also holds for the Dalitz decay of the
$a_1\to\pi\rho\to\pi \epm$. At higher bombarding energies,
however, direct processes like $\pi\rho\to \pi \epm$ might 
significantly contribute to the dilepton yield \cite{murray96a}.
\subsubsection{Dalitz decays of mesons}
\label{sectii-ii-i}
Decays of the type $A\to B \epm$ are not real multi body
decays (for which usually the name Dalitz is
used) but can be reduced to a decay into $B$ plus
a virtual photon (with an invariant mass $M$) 
and subsequent conversion of
the latter.
Thus the matrix element factorises
\cite{craigie78,landsberg85}
\be
|\M|^2=|\M(A\to B \gamma^\ast)|^2 \frac{1}{M^4}
|\M(\gamma^\ast\to\epm)|^2\; .
\ee
This reflects in the differential decay rate, which
can be 
written as a product of the conversion rate
of the virtual gamma \cite{kroll55}
and the decay width $\Gamma_{A\to B\gamma^{\ast}}$
of $A$ into a massive photon:
\be
\frac{d \Gamma_{A\to B\epm}}{dM}=
\frac{2\alpha}{3\pi M} \sqrt{1-\frac{4m_e^2}{M^2}}\,
\left(1+\frac{2m_e^2}{M^2}\right) \Gamma_{A\to
B\gamma^\ast}\; .
\label{eq_ddlzd2}
\ee
Here $m_e$ is the lepton mass. If one assumes
that $A$ decays isotropically in its restframe
the $A\to B \gamma^\ast$ width is given by 
\be
\Gamma_{A\to B\gamma^{\ast}}=\frac{|\vec p_{cm}|}{8\pi
m_A^2}|\M_{A\to B \gamma^\ast}|^2\; .
\label{eq_gamma2gam}
\ee
Here $|\vec p_{cm}|$ is the decay momentum
\be
|\vec
p_{cm}|=\frac{\lambda^{\frac{1}{2}}(m_A^2,m_B^2,M^2)}{2m_A}
\ee
with the kinematic function 
$\lambda(x^2,y^2,z^2)=(x^2-(y+z)^2)(x^2-(y-z)^2)$.

A remaining difficulty resides in the calculation of the
matrix element of the $A\to B\gamma^\ast$ decay.
The meson decays considered here are either of the type
a) pseudoscalar ($\pi^0,\eta,\eta'$) into a vector
particle ($\gamma$) plus a virtual photon or
b) vector meson ($\omega$) into a pseudoscalar meson plus
a virtual photon.
In both cases one gets \cite{kochp93a}
\be
|\M_{A\to B\gamma^\ast}|^2
=\frac{1}{2}|f_{AB}(M^2)|^2\lambda(m_A^2,m_B^2,M^2)\; .
\ee
The form factor $f_{AB}(M^2)$ is introduced to
account for the strong interaction part of the
vertex. It is common to normalise to the decay
width into real photons \cite{kochp93a} by dividing
\be
\Gamma_{A\to B \gamma}=\frac{(m_A^2-m_B^2)^3}{32
\pi m_A^3} |f_{AB}(0)|^2\; ,
\ee
which re-expresses the form factors to 
$F_{AB}(M^2)=\frac{f_{AB}(M^2)}{f_{AB}(0)}$.

The form factors can be obtained from the vector meson dominance
model (VMD). In the present calculations 
the following parametrisations are employed
\cite{landsberg85,ko96a}
\be
F_{\pi^0}(M^2)&=&1+b_{\pi^0} M^2\nonumber\\
F_{\eta}(M^2)&=&\left(1-\frac{M^2}{\Lambda_\eta^2}\right)^{-1}\nonumber\\
\left|F_\omega(M^2)\right|^2&=&
\frac{\Lambda_\omega^2(\Lambda_\omega^2+\gamma_\omega^2)}
{(\Lambda_\omega^2-M^2)^2+\Lambda_\omega^2\gamma_\omega^2}\nonumber\\
\left|F_{\eta'}(M^2)\right|^2&=&
\frac{\Lambda_{\eta'}^2(\Lambda_{\eta'}^2+\gamma_{\eta'}^2)}
{(\Lambda_{\eta'}^2-M^2)^2+\Lambda_{\eta'}^2\gamma_{\eta'}^2}
\label{eq_VMDform}
\ee
with $b_{\pi^0}=5.5\GeV^{-2}$, $\Lambda_\eta=0.72\GeV$,
$\Lambda_\omega=0.65\GeV$,
$\gamma_\omega=0.04\GeV$,
$\Lambda_{\eta'}=0.76\GeV$ and
$\gamma_\eta'=0.10\GeV$.
\subsubsection{Delta Dalitz decay}
\label{sectii-ii-ii}
The situation is more complicated for the Dalitz decay of
the
$\Delta(1232)$ resonance. To complete eqs. (\ref{eq_ddlzd2}) and
(\ref{eq_gamma2gam}), the matrix element for the process
$\Delta\to N \gamma^\ast$ has to be calculated. The
corresponding interaction vertex has been analysed by Jones
and Scadron \cite{jones73a} in the form
\be
\L_{int}=e \bar \Psi_\beta \Gamma^{\beta\mu} A_\mu \psi + h.c.\; ,
\label{eq_LDel}
\ee
where $\psi$, $\Psi_\beta$ and $A_\mu$ represent
the fields of the nucleon, of the delta and of the
photon, respectively.
The dominant magnetic dipole
transition yields the vertex function
\be
\label{eq_gammadel}
  \Gamma^{\beta\mu} = G_M(M^2 )
    \frac{-3(m_\Delta+m_N)}{2m_N((m_\Delta+m_N)^2-M^2)}
      \left (
        -m_\Delta \chi_1^{\beta\mu}\gamma_5
+\chi_2^{\beta\mu}\gamma_5
+\frac{1}{2}
   \chi_3^{\beta\mu}\gamma_5 \right)\; .
\ee
The choice of the $\chi$ bears some freedom if only
current conservation is ensured
($q_\beta\Gamma^{\beta\mu}=0$). 
Following ref. \cite{jones73a}, one may write
\be
\chi_1^{\beta\mu}& =  &
\(q^\beta \gamma^\mu-q\cdot\gamma g^{\beta\mu}\) \;,  \nonumber \\
\chi_2^{\beta\mu}& =  &
\(q^\beta P^\mu-q\cdot P g^{\beta\mu}\) \;,   \\
\chi_3^{\mu \beta}& =  &
\(q^\beta q^\mu-q^2 g^{\beta\mu}\) \;,\nonumber
\ee
with $P=1/2 (p_\Delta+p_N)$ and $q=p_\Delta-p_N$.                  
$G_M$ in eq. (\ref{eq_gammadel}) stands for the magnetic dipole
form factor. It can be fixed at $M=0$ to the decay into
a real photon, yielding $G_M(0)=3.0$.  
The $M$-dependence of the overall form factor is
subject to speculation: the
time-like electromagnetic form factors for baryons are
unknown in the kinematic region of interest.
In the present calculations 
any VMD-type form factors are omitted. This gives
a lower limit for the $\Delta$ Dalitz
contribution in the region of the vector meson poles.

Using eqs. (\ref{eq_LDel},\ref{eq_gammadel}), we can express the matrix
element via
\be
\hat\M_i&=&e \bar u_\beta(p_\Delta,s_\Delta)
\chi^{\beta\mu}_i\gamma_5 \epsilon_\mu(p_\gamma,s_\gamma)
u(p_N,s_N)\; .
\ee
The matrix element may be written as a linear
combination of the $\hat \M_i$ and $\hat \M_j^\ast$:
\be
\label{eq_delmat1}
|\M|^2=e^2G_M^2\frac{9\(m_\Delta+m_N\)^2}
 {4 m_N^2\((m_\Delta+m_N)^2-M^2\)^2}
 \sum_{i,j=1}^3 c_i\hat\M_i c_j\hat\M^\ast_j\; ,
\ee
where $c_1=-m_\Delta$, $c_2=1$ and
$c_3=1/2$.
The appearing traces have been calculated using the 
$Mathematica$ package HIP \cite{hsieh92}. One
then obtains:
\be
\label{eq_delmat2}
&& \sum_{i,j=1}^3 c_i\hat\M_i c_j\hat\M^\ast_j=
 \frac{1}{9}\((m_\Delta-m_N)^2-M^2\)\nonumber\\
&&\times\(7m_\Delta^4+14m_\Delta^2M^2+3M^4+
 8m_\Delta^3m_N+2m_\Delta^2m_N^2+6M^2m_N^2+
  3m_N^4\)\; .
\ee
Substituting eqn. (\ref{eq_delmat1}) and (\ref{eq_delmat2}) into 
eq. (\ref{eq_gamma2gam}), we get the 
$\Delta$ Dalitz decay width.

Fig. \ref{dalspectra} shows the differential mass
distributions of the Dalitz decay probabilities as 
implemented into the UrQMD model.
While the very low masses are dominated
by $\pi^0$ and $\eta$ decays, the $\omega$
and $\eta'$ decays are more important at higher masses.
The contributions of the $\Delta$ decays are
also shown for different masses of the $\Delta(1232)$. Heavy $\Delta$'s
naturally contribute more to the probabilities than lighter
ones.
In calculating the dilepton spectrum
these probabilities are multiplied with the yields of the
corresponding particle species.
At low energies the dominant
$\pi N\Delta$ system \cite{bass95c} can push 
the $\Delta$ contributions over those of the heavy
mesons. On the other hand, the small yield of the
$\eta'$ will make it invisible, due to the tremendous
background of other sources in the relevant mass region.
\subsubsection{Direct decays of vector mesons}
\label{sectii-ii-iii}
The decay width of the electromagnetic two-body decays of the vector
mesons are assumed (similar to ref. \cite{ko96a}) 
to be of the form
\be
\label{eq_decvec}
\Gamma_{V\to \epm}(M)=\frac{c_V}{M^3}
\sqrt{1-\frac{4m_e^2}{M^2}}\,
\left(1+\frac{2m_e^2}{M^2}\right)
\Theta(M-2m_e) 
\; .
\label{eq_gammaee}
\ee
The constants $c_V$ are fitted to yield the
vacuum widths given in ref. \cite{PDG96} at the
resonance poles. One obtains
$c_\rho=3.079\times 10^{-6}\GeV^4$,
$c_\omega=0.287\times 10^{-6}\GeV^4$ and 
$c_\phi=1.450\times 10^{-6}\GeV^4$.

It is assumed that vector mesons can radiate off dileptons
continuously \cite{heinz92a}. The dilepton mass distribution
is given by the time-integral over the
mass distributions of mesons plus an additional  term 
which accounts for the decays after some typical freeze-out time $t_f$
(the last term can be dropped if $t_f\to\infty$)
\be
\frac{dN_{ee}}{dM}=
\int_0^{t_f} dt
\frac{dN_V(t)}{dM}\cdot\Gamma_{V\to ee}(M)+
\frac{dN_V(t_f)}{dM}\cdot\frac{\Gamma_{V\to
ee}(M)}{\Gamma_{V,tot}(t_f)}\; .
\label{eq_dndmshine}
\ee
If absorption is negligible, this
approach is equivalent to the
method of adding one dilepton (with appropriate
normalisation) at each
decay vertex. However, this ''shining'' method
gives a better sampling of the density probed by the hadron
and thus a better statistics for density dependent spectral
functions.
In addition, there is generally a complicated
time dependence of the total
hadron width $\Gamma_{tot}(t,\vec r)$, caused by the
in-medium modifications of the quasiparticle widths.
The approach 
eq. (\ref{eq_dndmshine})
has the advantage to be explicitely independent of
the in-medium width, if $t_f$ is reasonably high,
so that the total width takes on it's vacuum properties.
\subsubsection{Incoherent bremsstrahlung}
\label{sectii-ii-iv}
In the soft photon approximation (SPA) the
cross section for real photons with
four-momentum $q^\mu$ in
collisions $a+b\to X$ can be expressed as
\be
q_0\frac{d\sigma^{\gamma X}_{ab}}{d^3 q}=
\frac{\alpha}{4\pi^2}\left|\epsilon\cdot J\right|^2 
d\sigma_{ab}^X\; .
\label{eq_sigab2Xg}
\ee
$d\sigma_{ab}^X$ denotes the differential cross section
of the (strong) $ab$ interaction without any photon in the
final state.
$\epsilon$ is the polarisation vector of the
photon. The current $J^\mu$ is given by
\be
J^\mu(q)=-Q_a\frac{p^\mu_a}{p_a\cdot q}
-Q_b\frac{p^\mu_b}{p_b\cdot q}
+\sum_{i=1}^{X}Q_i\frac{p^\mu_i}{p_i\cdot q}\; ,
\ee
where the $Q$'s and $p$'s denote the charges and momenta
of the corresponding particles.

The SPA is justified only for $M,q_0\to 0$. To
extrapolate to the case of hard and massive virtual
photons, a phase 
space correction can be applied by multiplying
the cross section
with the ratio of the phase space integrals with/without a
virtual photon \cite{gale89b}. Similarly to eq.
(\ref{eq_ddlzd2}) one gets
\be
\frac{d\sigma^{\epm X}_{ab}}{d^3 q dM}=
\frac{\alpha^2}{6\pi^3}\sqrt{1-\frac{4m_e^2}{M^2}}\,
\left(1+\frac{2m_e^2}{M^2}\right)
\left|\epsilon\cdot J\right|^2 
\frac{R_n(\bar s)}{R_n(s)}\frac{d\sigma_{ab}^X}{q_0 M}\; .
\label{eq_xx2brems}
\ee
Here $R_n$ is defined as
\be
R_n(s)=\int d\Phi_n(s,p_1\ldots p_n)\; ,
\ee
where $d\Phi_n$ is the volume element of the
$n$-dimensional Lorentz invariant phase space and $\bar
s$ is the squared effective energy
of the system after the emission of the
$\gamma^\ast$,
\be
\bar s= s+M^2-2\sqrt{s}q_0\; .
\ee
The correction factor for two outgoing particles reads
\be
\frac{R_2(\bar s,m_a^2,m_b^2)}{R_2(s,m_a^2,m_b^2)}=
\frac{\lambda^{1/2}(\bar s,m_a^2,m_b^2)}{\lambda^{1/2}(s,m_a^2,m_b^2)}
\frac{s}{\bar s}\; .
\ee

Eq. (\ref{eq_xx2brems}) is a general
expression for the bremsstrahlung 
dilepton production in the SPA.
In the case of proton-neutron
bremsstrahlung ($pn\to p'n'\gamma^\ast$), the
current is given by $J^\mu=p_{p'}^\mu/q\cdot p_{p'}-p_{p}^\mu/q\cdot
p_{p}$. As a result we have
\be
\left|\epsilon\cdot J\right|^2
=\frac{-t}{q_0^2 m_p^2}\; ,
\ee
with the Mandelstam variable $t=(p_p-p_{p'})^2$. Then
eq. (\ref{eq_xx2brems}) takes the 
form 
\be
\frac{d\sigma^{pn\epm}_{pn}}{d^3 q dM}=
\frac{\alpha^2}{6\pi^3}\sqrt{1-\frac{4m_e^2}{M^2}}\,
\left(1+\frac{2m_e^2}{M^2}\right)
\frac{\bar \sigma}{q_0^3 M} 
\frac{R_2(\bar s)}{R_2(s)}\; .
\label{eq_pn2brems}
\ee
Here $\bar \sigma$ is the
momentum-transfer weighted $pn$ elastic cross section
\be
\bar \sigma=\int_{-(s-4m_p^2)}^0 
\(\frac{-t}{m_p^2}\)\frac{d\sigma}{dt}dt\; .
\label{sigmabar}
\ee
A proper parametrisation
of $\bar \sigma$ is rather important \cite{winckelmann93a}.
A systematic overestimation of the differential bremsstrahlung
production results if the asymmetry of the momentum-transfer weighted 
cross section is not included.
In the present work, a parametrisation similar to
that of ref.
\cite{winckelmann93a} is used.
It consists of a
symmetric low energy part and an asymmetric
high energy component.

Fig. \ref{bremsdm} shows the cross section for
$pn\to pn+\epm$ bremsstrahlung
at different bombarding energies. Note that at low
masses, 
the cross section is practically not sensitive to the
incident energy. As a result, the $pn$ bremsstrahlung
is relatively unimportant
as resonance
channels become dominant at $E_{kin}\sim 1.5\GeV$.

Bremsstrahlung from $pp$ collisions
is expected to be small at low
energies due to destructive interferences.
However, at the highest energies considered here, the $pp$
bremsstrahlung contribution  may be comparable to the $pn$ 
bremsstrahlung yield \cite{haglin94a}. On the
other hand, the total contribution of 
bremsstrahlung is found to be negligible for all practical
purposes at those energies (see sect. \ref{sectiii-ii}).
Note that the SPA already gives 
an upper estimate of the expected dilepton yield \cite{lichard95a}.
\section{Dilepton cross sections in elementary
systems}
\label{sectiii}
Before one can investigate the
dilepton production in
nucleus-nucleus reactions, one should first check
the model in
elementary $pp$ and $pd$ collisions.
Recently, the DLS collaboration
published a systematic study on dielectron cross
sections in light systems for beam kinetic energies 
$\T$ from 1 to 5~GeV. The corresponding 
centre-of-mass energies range from
below the $\eta$ production threshold up to
3.6~GeV, where 
phase space is wide open
for the abundant production of a large variety of resonances. 

The acceptance of the DLS does not cover
the entire phase space. Thus the calculated dilepton
pairs are corrected for the limited
spectrometer acceptance region (DLS-filter~V4.1).
The finite mass resolution of $\Delta
M/M\approx 0.1$ is incorporated by folding the calculated spectra
with a Gaussian of width $\Delta M$. 
The acceptance correction strongly influences the low-mass
spectra while
the mass resolution smoothing
affects the shape of the spectra at higher masses.
\subsection{pp}
\label{sectiii-i}
The resulting dilepton mass spectra for $pp$ collisions are shown in
fig. \ref{dlspp}. 

Only $0.46\GeV$ of c.m. energy are
available for particle production at
$\T=1.04\GeV$. Thus,
only pion- and $\Delta$-Dalitz decays
contribute in this case.
The model satifactorily reproduces the data at low masses,
but 
underestimates them around
$M_{\epm}=0.4\GeV$.
The disagreement could be caused by the
abovementioned uncertainties in calculating
the electromagnetic form factor of the $\Delta N\gamma$
system.
The first
generation of DLS data was
incompatible to free form factors \cite{wolf90a,schaefer94a}.
This situation
is now unclear, because the data of the second run
tends to exceed the first.
However, limited statistics and
the better agreement at higher beam energies
precludes a definite conclusion
on the origin of the ''enhancement'' at
$\T=1.04\GeV$.

At $\T=1.27 - 1.85\GeV$ the model
explains the data with growing
influence of meson decays. One can see from fig.
\ref{dlspp} that first
the $\eta$-Dalitz decay and then the direct $\rho$ and
$\omega$ decay become more important. In the UrQMD
calculation of $pp$, most of the $\rho^0$
($\approx 99\%$)
are not created in $\pi^+\pi^-$ collisions, but 
result from heavy baryon decays. The limited
phase space at low c.m. energies \cite{winckelmann95a} 
and the strong ($M^{-3}$) mass dependence of the dilepton decay
widths (\ref{eq_decvec})
are responsible for the deformation of the $\rho$
spectrum towards low masses.
The cross section
for the Dalitz production through $\Delta$ decays
remains rather constant at low masses, with
increasing energy, but 
increases towards higher masses.
Indeed, it is known
from $pp\to n\Delta^{++}$ reactions, that heavier
$\Delta$'s become more important with increasing beam
energy due to the mass dependence of the  decay widths
\cite{dmitriev86a}.
In fact, there is a good agreement at
intermediate beam energies (disregarding the
data point at $M_{\epm}=200\MeV$) which would be
worsened if the electromagnetic $\Delta N \gamma$ 
form factor would be included,
in particular at $\T\ge 1.6\GeV$.

The model does not reproduce the shape of
the measured spectrum at $\T=2.1\GeV$ and
4.9~GeV.
Around $M=0.6\GeV$ the
ratio between the $pp$ and $pd$ data decreases
for the two highest beam energies.
If this gets confirmed, it might be explainable by 
some strong contribution of
$pp$ bremsstrahlung, which was not considered here.
\subsection{pd}
\label{sectiii-ii}
$pd$ calculations are shown in fig. \ref{dlspd}.
The calculated events are minimum-bias triggered at a
maximum impact parameter of $b=1.6\fm$, i.e.
a geometric cross section equal to the measured
one of about 80\mb.

An important difference to the $pp$ system is
due to the internal motion of bound nucleons.
This motion smears out the production thresholds.
Consequently, one observes 
subthreshold contributions from 
$\eta$ and $\rho$ mesons already at $\T=1.04\GeV$.

Unlike in $pp$ collisions, the 
proton-deuteron data
overestimate the model results
in the mass region dominated by the $\eta$ decay.
This indicates an asymmetry in the $pp$ and
$pn$ production cross sections as predicted from
one-boson exchange models \cite{vetter91a}.
Indeed, it was
measured that just above the threshold, the ratio of
$pd$ to $pp$ induced $\eta$ cross sections is
much larger than two \cite{chiavassa94a}. However, due to
the Fermi motion in
the deuteron, it is not straight forward
to extract $pn$ cross sections 
from $pp$ and $pd$ measurements. On the other
hand, direct measurements
of the $\eta$ production in $pn$ are still not
available. We have therefore evaluated
the ratio
$R_\eta=\sigma(pd\to\eta X)/\sigma(pp\to\eta X)$ without any explicit
modification of the $pn$ cross section to estimate
the influence of Fermi motion.
At $\T=1.27\GeV$, close to threshold, 
we get the large value $R_\eta=17\pm 3$ while
above $\T=1.61\GeV$,
$R_\eta$ remains practically constant,
between 2.8 and 2.3.

However, the large value of $R_\eta$ is not enough to
explain 
the integrated total dilepton ratios. This is shown in
fig.~\ref{pd2pp} where 
\be
R=\frac{\int_{0.15~GeV}^{M_{max}}\(\frac{d\sigma_{pd}}{dM}\)dM}
	{\int_{0.15~GeV}^{M_{max}}\(\frac{d\sigma_{pp}}{dM}\)dM}
\ee
is plotted vs. the incident energy.
The predicted dilepton ratios are rather low at energies 
smaller than 2~GeV. A similar behaviour is also found
in BUU calculations \cite{titov95a}, which include $pp$
bremsstrahlung. This indicates that there is a
common feature of known transport
models with ''conventional'' sources to underestimate
the recent DLS nucleus-data at intermediate 
dilepton masses \cite{porter97a}.  At
$\T=1.27\GeV$ the model predicts only $R=5.2$. This value is lower
than the corresponding one of $R_\eta$, because
the $\eta$ is yet only a marginal dilepton source at this
energy.

Within the statistical uncertainties the mass-differential 
dilepton cross sections for $pd$ can be explained
by the model for the two highest beam energies.
Note that $pn$ bremsstrahlung is relatively unimportant.
\section{Dileptons in heavy ion collisions}
\label{sectiv}
Fig. \ref{dlsaa} compares the UrQMD predictions
with DLS data for various nucleus-nucleus
reactions \cite{porter97a}.
At first glance the
examined systems all exhibit a qualitatively similar
behaviour in three distinct mass regions:

In the lowest mass region (up to 150\MeV) 
the spectrum
is dominated by the pionic Dalitz decay. There
are also strong effects of the acceptance filter,
which significantly suppresses the low-mass
yield (compare Fig. \ref{dalspectra} for the
shape of uncorrected Dalitz spectra).
Other sources are of little 
importance in this mass region. However,
near $M=150\MeV$, the $\Delta$ Dalitz decay
becomes more important.

At intermediate masses
the data show a noticeable enhancement of the dilepton yield as
compared to the model calculations.
There is a considerable 
confusion about these recent data
because the new yields strongly exceed earlier 
published measurements \cite{roche89a}.
The latter have been revised by the DLS
collaboration due to large, previously uncorrected trigger
inefficiencies.
However, the present calculations and, also, results of other models
e.g. \cite{wolf90a,xiong90a,bratko96b,bratko97a},
are closer to the 
earlier measurements in
the considered mass range (see fig.~\ref{ca1ca}).

The theoretical spectrum at $M>600\MeV$ is dominated by
direct $\rho^0$ and $\omega$ decays.
The model cross sections for dilepton production via
$\rho$ mesons nicely account for
the data in this region.
The two highest data points in He-Ca suffer from
lack of statistics.
For all nuclear systems,
only about 50\% of all $\rho$ mesons  stem from 
$\pi^+\pi^-$ annihilations, the other 50\% are produced
in decays \cite{winckelmann95a}.

Turning back to the intermediate region, one sees
a similar dilepton enhancement as in the $pd$
data. There one may partially attribute the
enhancement  to the high
$pn\to\eta X$ cross section. Therefore, it seems
reasonable to reproduce the data
by artifically enhancing the $\eta$
yield. We rescaled the $\eta$ yield by a factor 
$f_\eta$ and found good agreement both in shape and
absolute yield with $f_\eta\approx 10$ for C+C and $\approx
20$ for Ca+Ca. 
However, this is not supported by TAPS data
which found the $4\pi$ extrapolated $\eta$ cross section
in Ca+Ca at 1~GeV
to be about 20~mb \cite{berg94a}.
Our rescaled cross section corresponds to
$\approx$120~mb. A dropping $\eta$ mass has been
introduced in ref. \cite{bratko97a} to explain the DLS
data. But these studies also found the increased
$\eta$ cross sections to be incompatible to the TAPS
measurements.

We note that the inclusion of a
density-dependent $\Delta N \gamma$ form factor 
for the $\Delta$ Dalitz decay, as
discussed in ref. \cite{wolf90a}, can give a 
sizeable enhancement of the calculated diletpon yield at
intermediate masses $0.2\GeV<M<0.6\GeV$.
However, this is ruled out as an explanation, because
the fair agreement of the calculation to the data in the
high-mass region $M>600\MeV$ would be destroyed by a strong
overestimation.

Due to $\sigma-\omega$ mixing via an $NN^{-1}$ loop
, the $2\pi$ decay channel of
the $\omega$ might be significantly enhanced in
nuclear matter \cite{wolf97a}. To get an estimate for
possible effects on intermediate mass lepton
pairs, we have increased
$\Gamma_{\omega\to\pi\pi}$ by a factor of 500 to
approximately 100~MeV. The solid curves in fig.~\ref{camed}
show the result of this calculation for the Ca+Ca system.
As can be seen, the mass distribution of dilepton 
pairs from direct $\omega$ decays becomes very
broad. Nevertheless, this contribution has almost no effect
on the total dilepton yield and does not suffice
to explain the measured enhancement.

One approach to describe the
density dependence of vector meson masses is the
linear scaling law of Hatsuda and Lee \cite{hatsuda92a}:
\be
m^\ast_V=m_V(1-0.18\varrho/\varrho_0)\; ,
\label{eq_hl}
\ee
where
$\varrho_{(0)}$ is the local (ground state) density and $m^{(\ast)}_V$
is the (modified) vector meson mass. Although more sophisticated
calculations predict a more complex behaviour of
the vector meson spectral functions
\cite{rapp97a}, the scaling
law seems to be reasonable for the
effective masses. To check this idea using the
UrQMD model,
the poles of the produced quasi particle vector
mesons have been shifted according to
(\ref{eq_hl}). One obtains the dashed curves in 
fig.~\ref{camed}. It is found that the
dropping mass scenario -- and also more complex in-medium
spectral functions (see ref. \cite{bratko97a}) -- cannot
account for the new DLS data.
\section{Summary}
\label{sectv}
Dilepton production has been studied within the microscopic non-equilibrium
transport model UrQMD. The production
mechanisms have been critically revisited. We
have compared the model with the DLS data
for $pp$ and $pd$ collisions at different
beam energies. Resonance decays
into dileptons were found to be able to
explain the low energy $pp$ data.

The UrQMD model predictions for the $pd$ system are below the new DLS
data for masses 0.3-0.6~GeV.
A similar, but even stronger underestimation takes place in
collisions of heavier nuclei. The
present paper
points out that the large enhancement in the data as compared to
model calculations cannot be accounted for 
by two distinct hypothesises on the in-medium modifications of 
vector mesons.
An enhancement of the 
$\eta$ production in $pn$ collisions is able to reproduce 
the yield and shape of the AA spectra. However, huge $\eta$ production
cross sections would be required, in contrast to TAPS data.

\section*{Acknowledgements}
C.~E. wants to thank A.~Dumitru, G.~Mao, L.~Satarov
and L.~Winckelmann for valuable
discussions and H.~S.~Matis for providing the DLS
acceptance filter.
 

\section*{Figure Captions}
\begin{figure}[htbp]
  \refstepcounter{figcounter}
  \vspace{0cm}
  \hskip -1cm\psfig{file=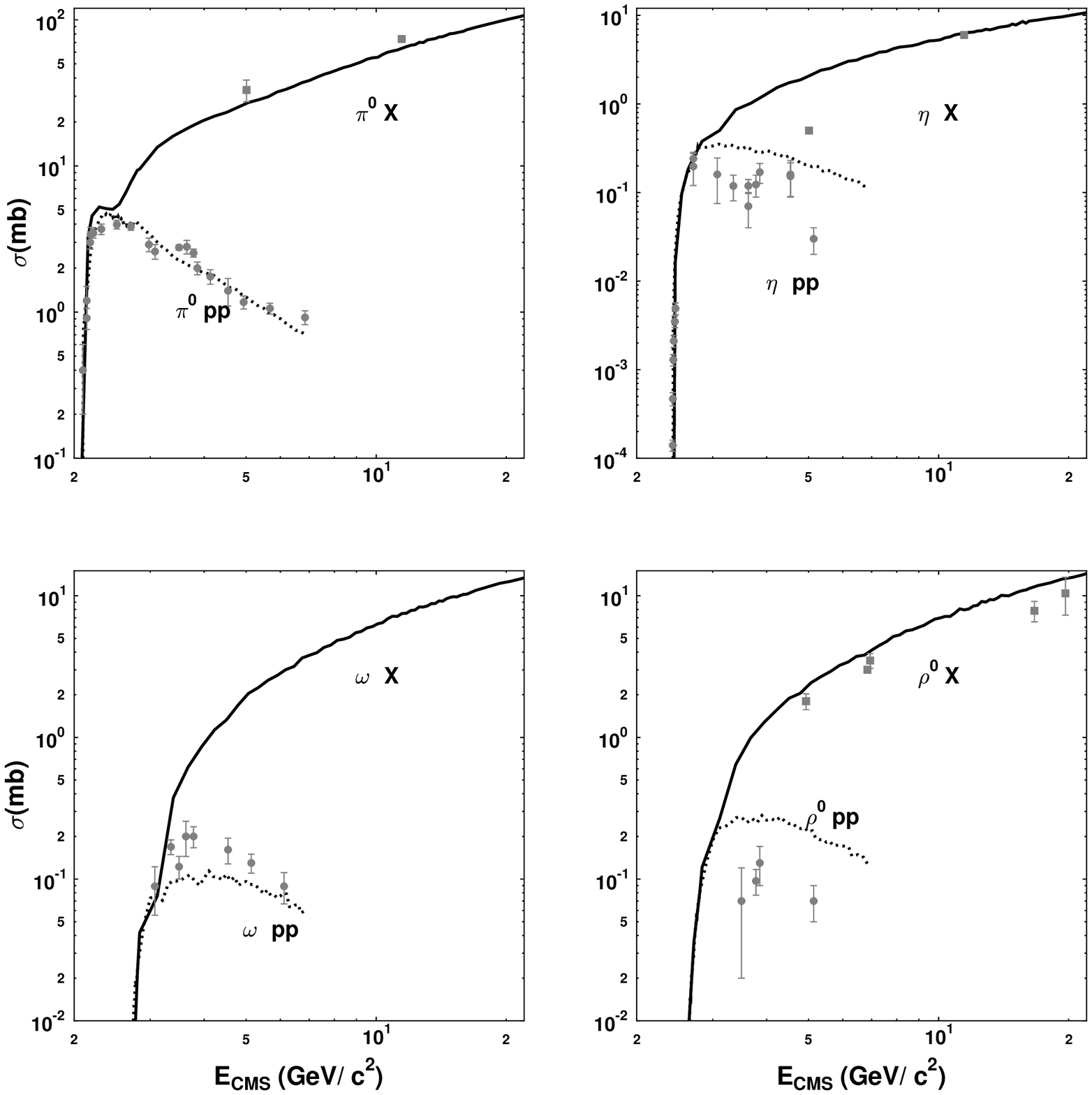,width=17cm,height=17cm}
  \vspace{-5cm}
  \label{ppc}
\end{figure}

\begin{figure}[htbp]
  \refstepcounter{figcounter}
  \vspace{0cm}
  \hskip 1cm\psfig{file=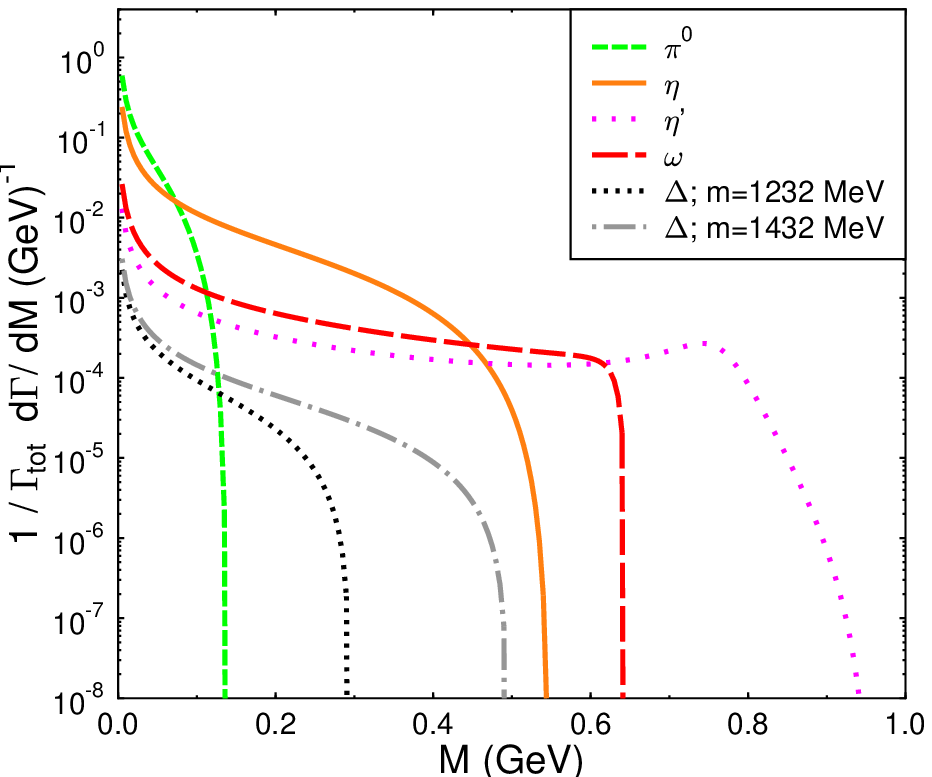,width=12cm,height=9cm}
  \label{dalspectra}
\end{figure}

\begin{figure}[htbp]
  \refstepcounter{figcounter}
  \vspace{0cm}
  \hskip 1cm\psfig{file=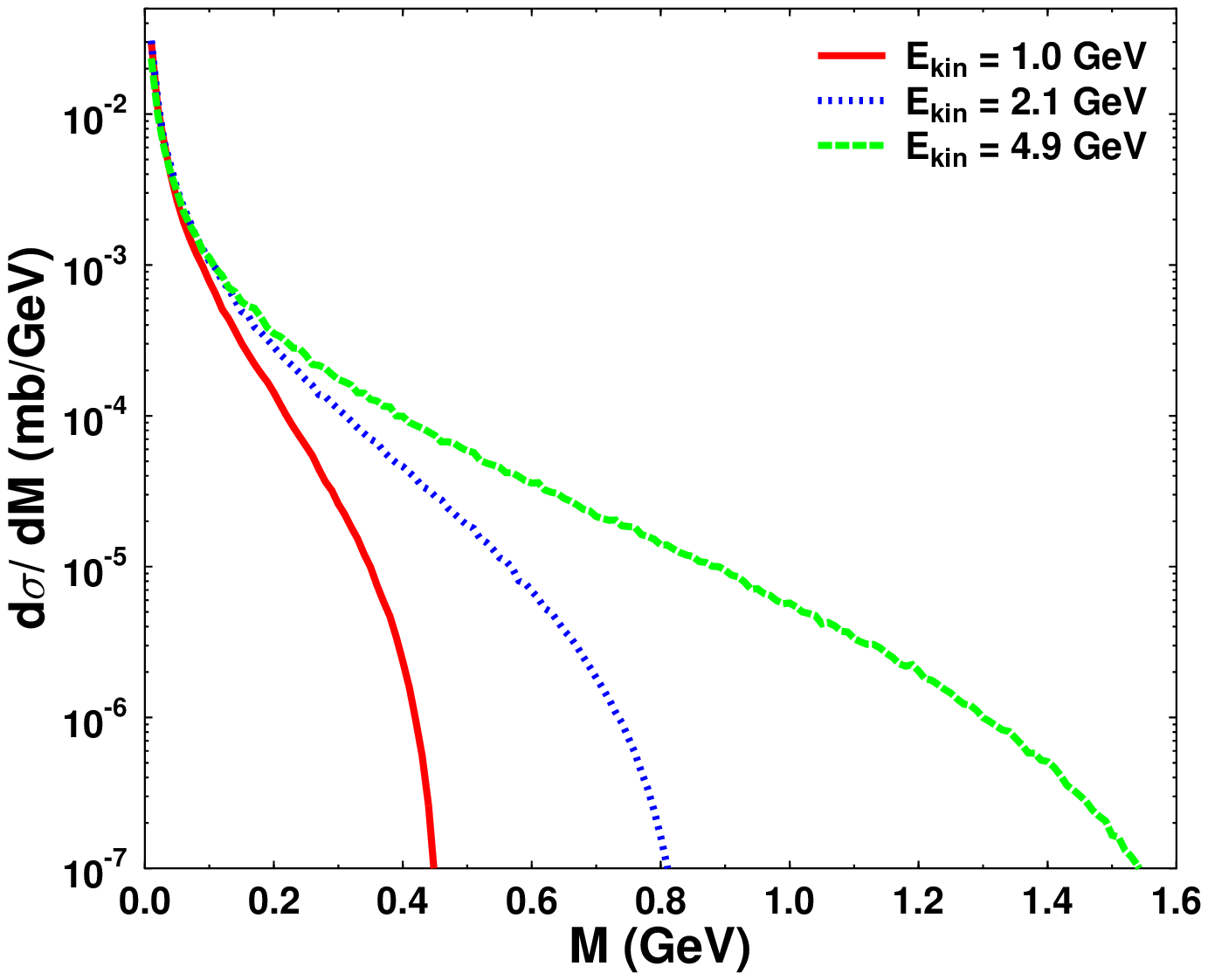,width=12cm,height=9cm}
  \vspace{-5cm}
  \label{bremsdm}
\end{figure}

\begin{figure}[htbp]
  \refstepcounter{figcounter}
  \vspace{-2cm}
  \hskip -1cm\psfig{file=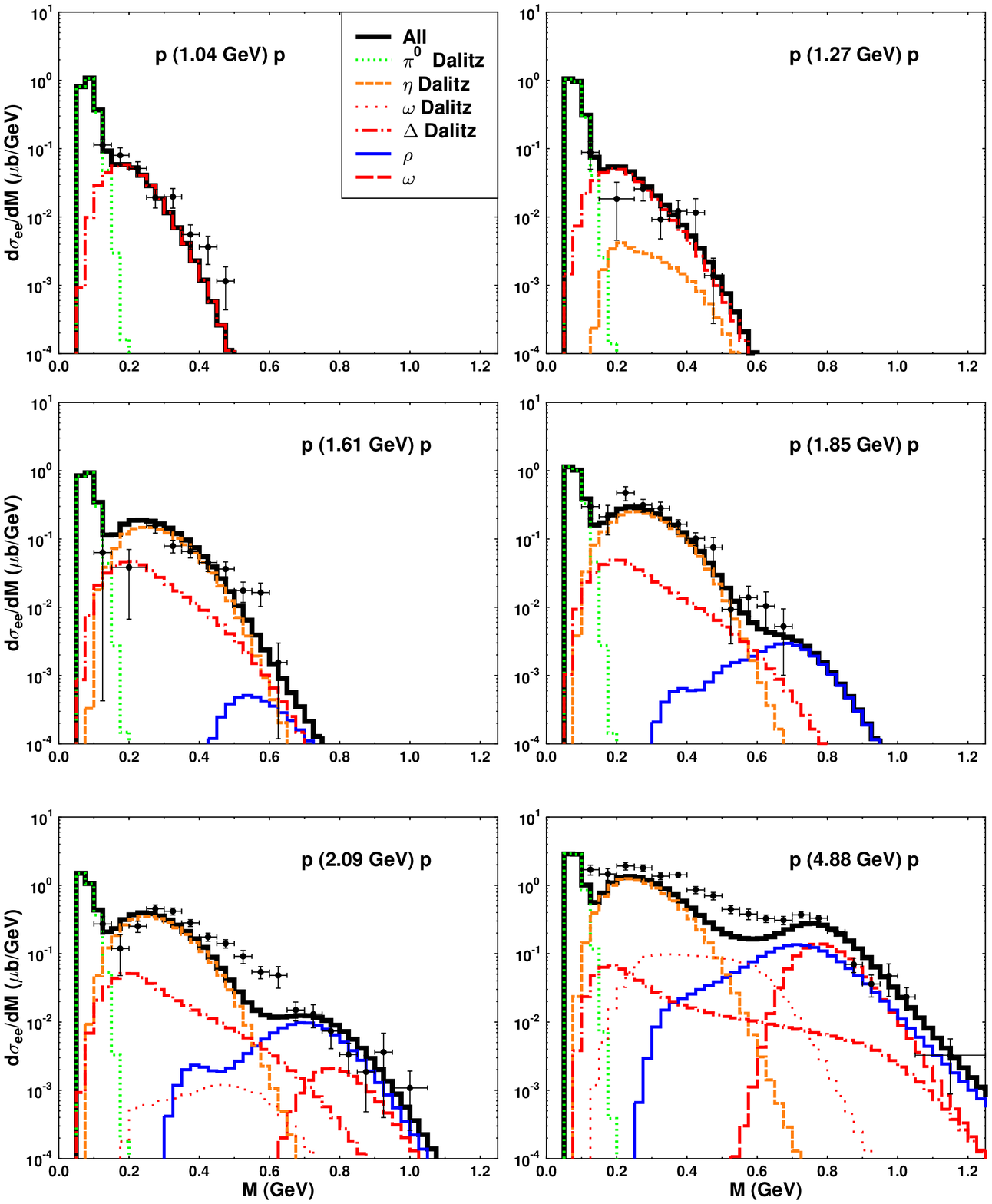,width=16cm,height=22cm}
  \vspace{-6cm}
  \label{dlspp}
\end{figure}

\begin{figure}[htbp]
  \refstepcounter{figcounter}
  \vspace{-2cm}
  \hskip -1cm\psfig{file=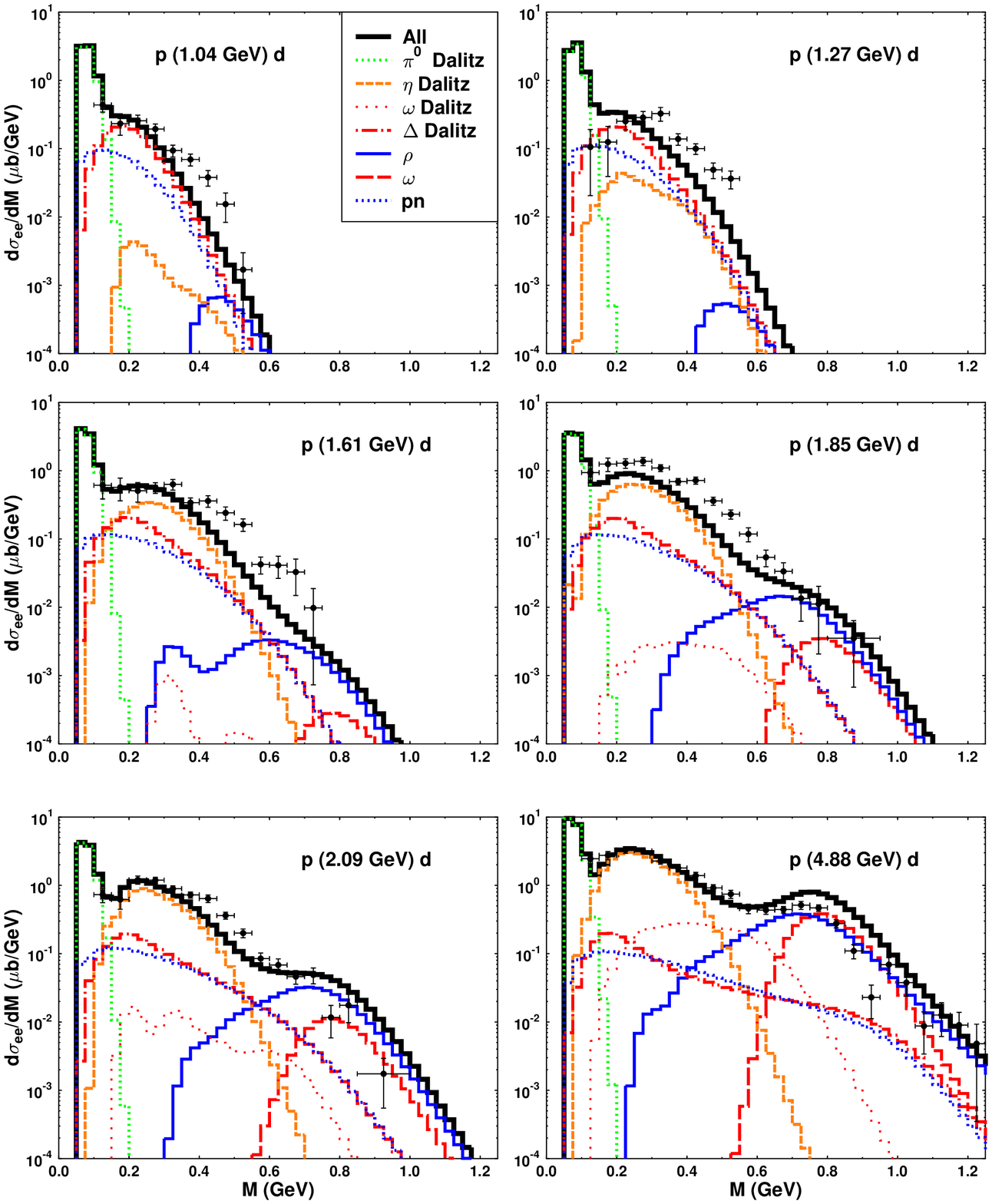,width=16cm,height=22cm}
  \vspace{-6cm}
  \label{dlspd}
\end{figure}

\begin{figure}[htbp]
  \refstepcounter{figcounter}
  \vspace{0cm}
  \hskip 1cm\psfig{file=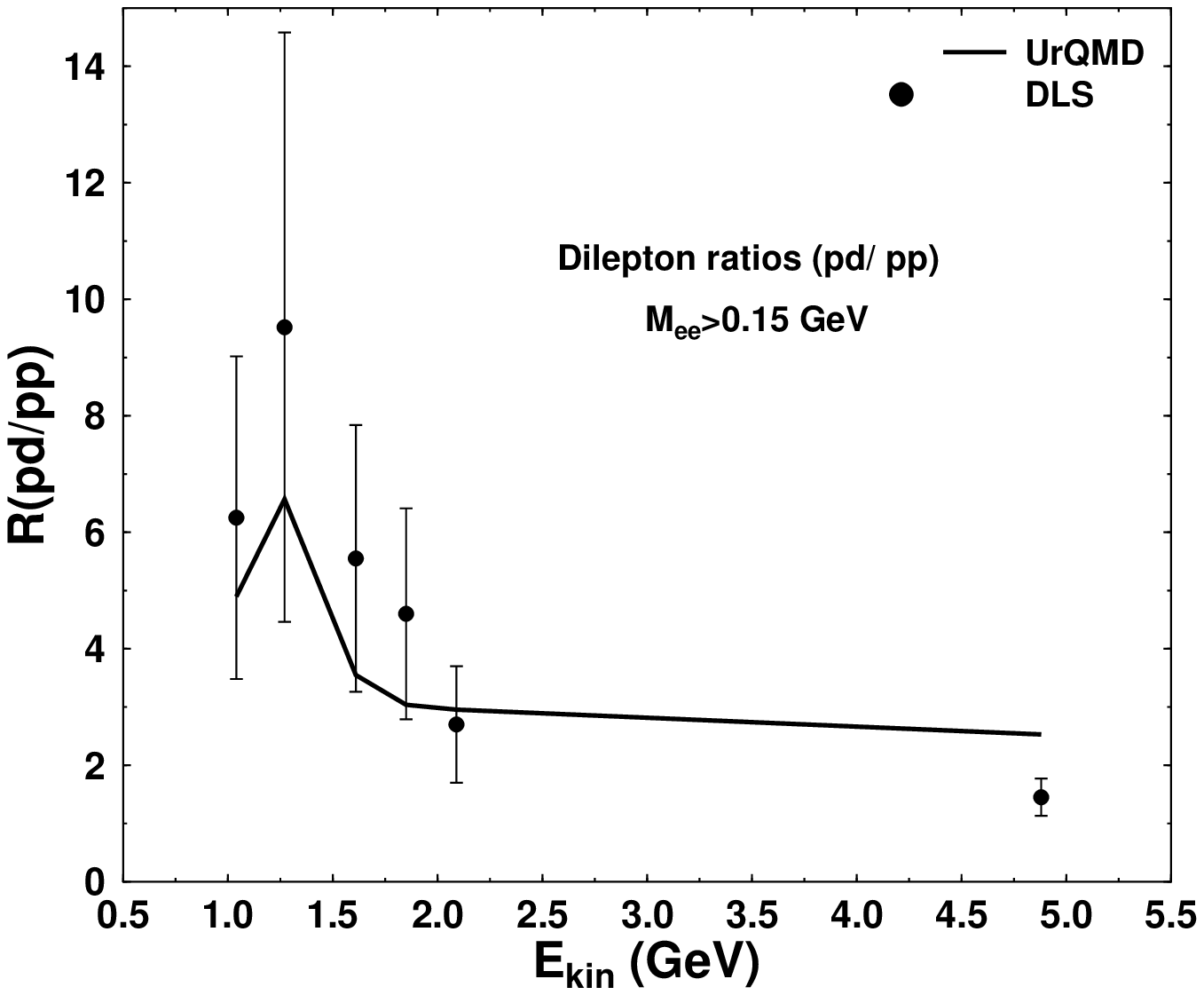,width=12cm,height=9cm}
  \label{pd2pp}
\end{figure}

\begin{figure}[htbp]
  \refstepcounter{figcounter}
  \vspace{0cm}
  \hskip -1cm\psfig{file=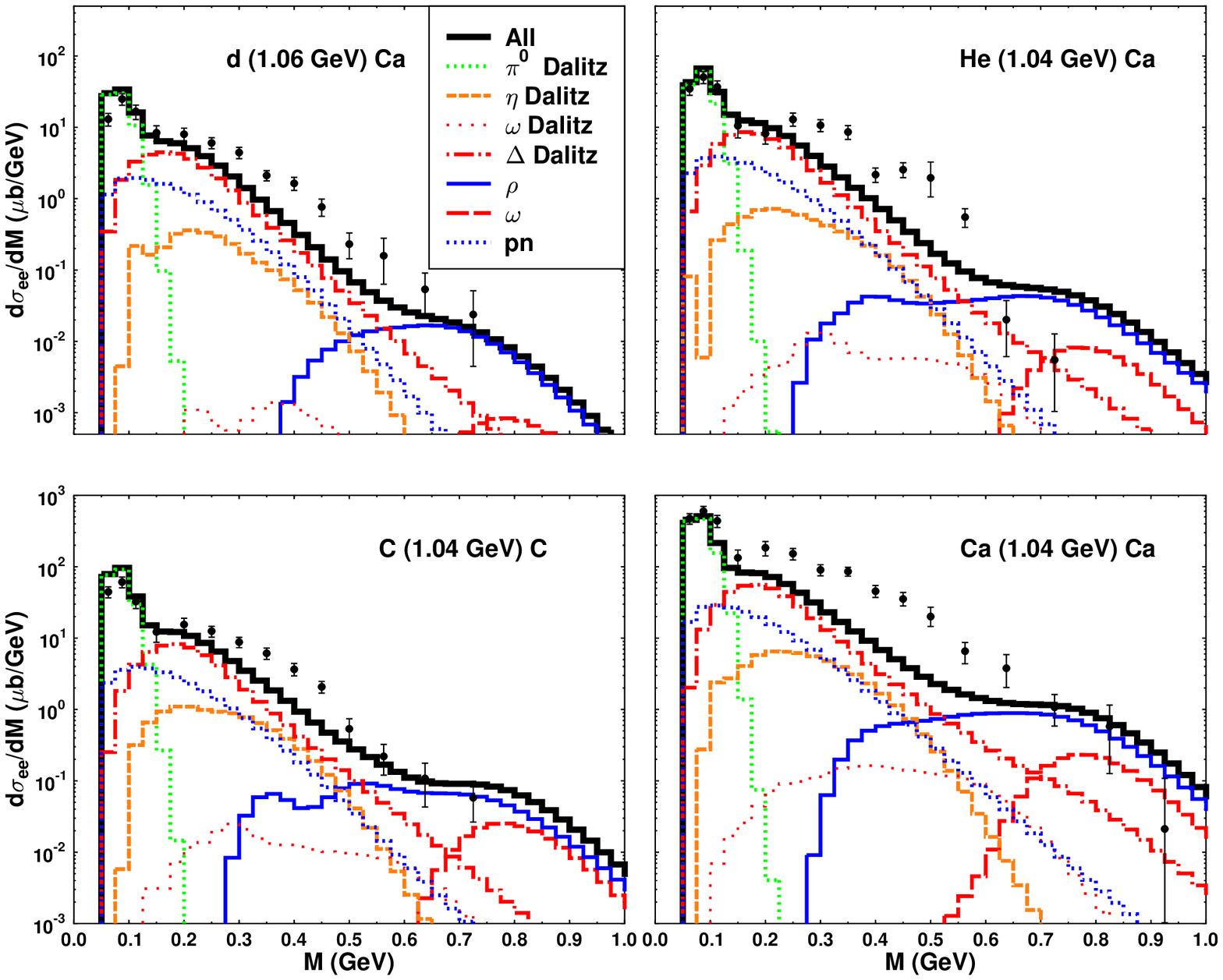,width=17cm,height=19cm,angle=-90}
  \vspace{-5cm}
  \label{dlsaa}
\end{figure}

\begin{figure}[htbp]
  \refstepcounter{figcounter}
  \vspace{0cm}
  \hskip 1cm\psfig{file=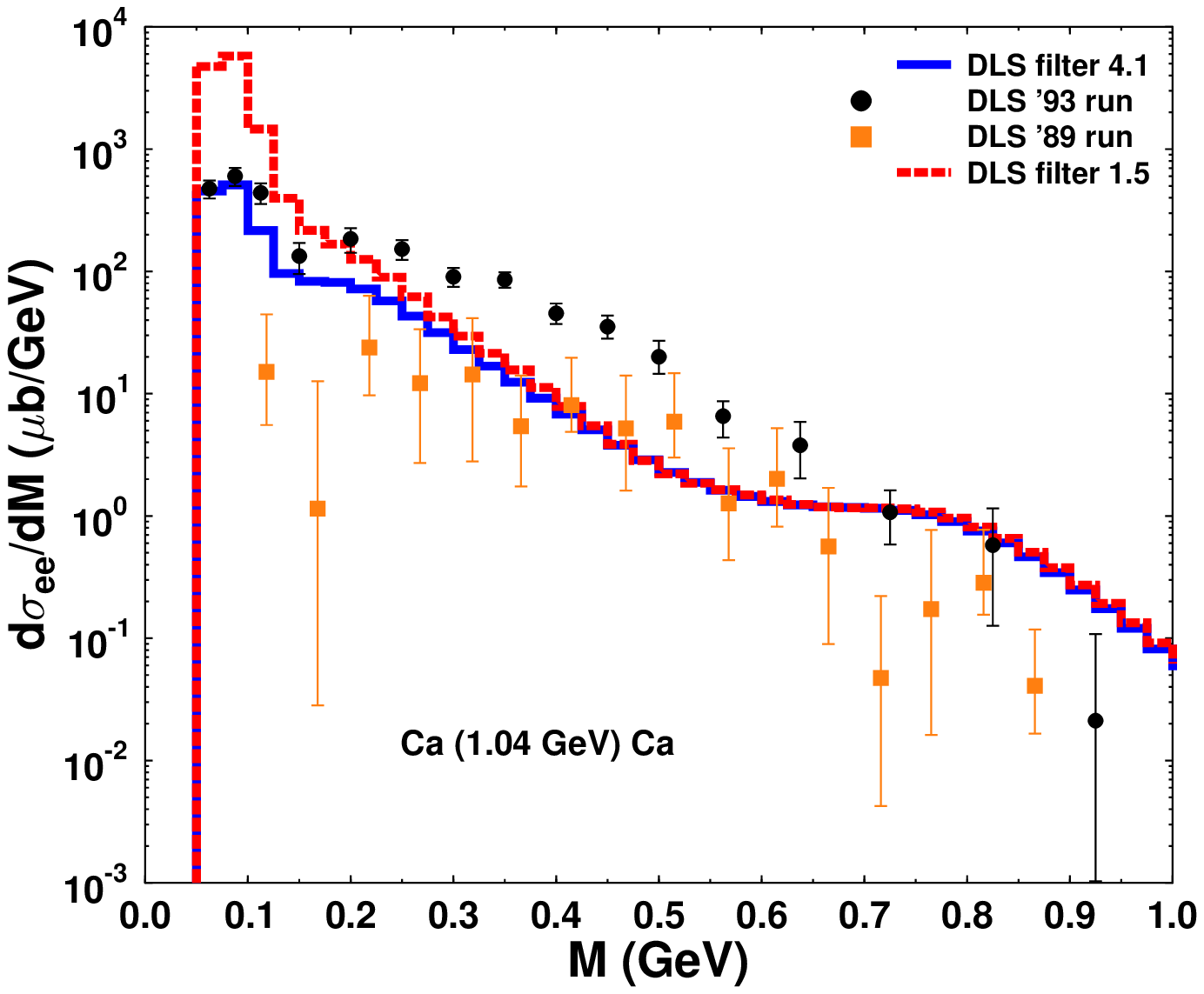,width=12cm,height=9cm}
  \label{ca1ca}
\end{figure}

\begin{figure}[htbp]
  \refstepcounter{figcounter}
  \vspace{0cm}
  \hskip 1cm\psfig{file=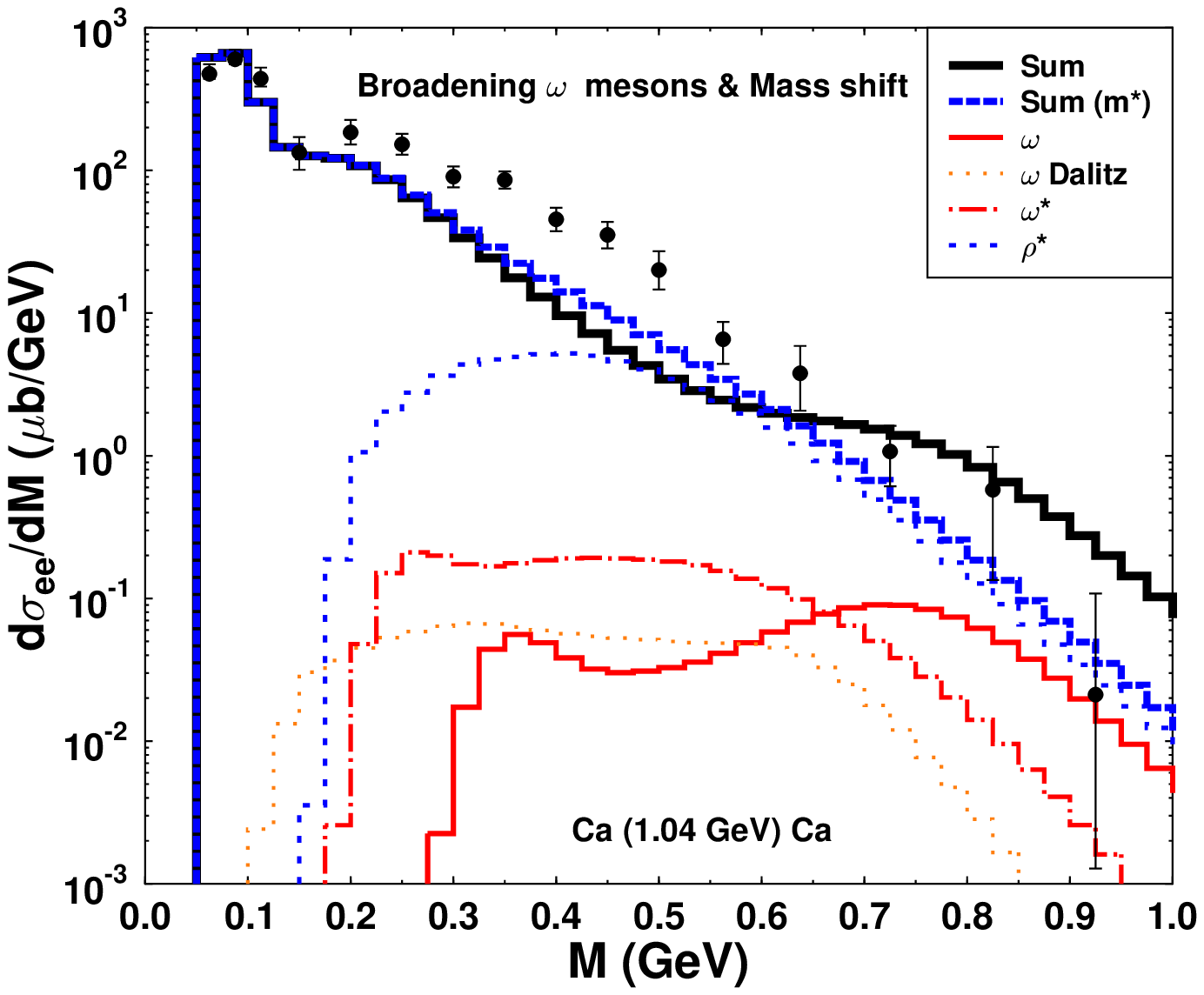,width=12cm,height=9cm}
  \vspace{-5cm}
  \label{camed}
\end{figure}

 \begin{description}
 \item[ FIG. \ref{ppc} ] Cross sections for neutral
meson production in $pp$ collisions. Calculations
are shown
for the exclusive and inclusive production of $\pi^0$,
$\eta$, $\omega$ and $\rho^0$ mesons in comparison to
avaliable data \cite{flaminio84a,calen96a}.
 \item[ FIG. \ref{dalspectra} ] Differential probability
distributions for Dalitz decays. The mass-differential
branching ratios for decays of $\pi^0$ (dashed), $\eta$
(solid), $\omega$ (long-dashed) and $\eta'$ (dotted) mesons can be
seen. Also shown is the parametrisation for Dalitz decays
of the $\Delta(1232)$ resonances at
pole mass (dotted) and of a heavy $\Delta(1232)$
with mass $m_\Delta=1.432\GeV$ (dash-dotted).
 \item[ FIG. \ref{bremsdm} ] Differential cross section for the
production of a bremsstrahlungs pair in elastic $pn$ collisions
for beam kinetic energies $\T=$1.04 (solid), 2.09
(dotted) and  4.88~GeV (dash-dotted).
 \item[ FIG. \ref{dlspp} ] The $\epm$ mass spectra for $pp$
reactions at six kinetic energies. 
The upper solid curve is the sum of all contributions.
One can see the contributions of $\pi^0$ (dotted),
$\Delta$ (dash-dotted), $\eta$ (dashed) and $\omega$ (dotted)
Dalitz decays as well as of the direct $\rho^0$
(solid) and $\omega$ (thin-solid) dilepton decays.
Every single dilepton has passed the 
DLS Filter 4.1 and a mass resolution of 10\%
is adopted. The data are from \cite{wilson97a}.
 \item[ FIG. \ref{dlspd} ] Acceptance-corrected mass spectra for
proton + deuterium. The individual sources are
marked are as in fig. \ref{dlspp}, but there is an
additional contribution from $pn$ bremsstrahlung
(thick-dotted).
 \item[ FIG. \ref{pd2pp} ] Ratio of the integrated cross
section for $pd$ to $pp$ reactions as a function of the
beam energy. The solid line shows the UrQMD result with
the DLS filter and resolution. The experimental data are taken
from ref.~\cite{wilson97a}. Only the larger
systematic error bars are displayed.
 \item[ FIG. \ref{dlsaa} ] Model calculations
for dilepton spectra from
nucleus-nucleus collisions at beam energies of
about 1~GeV in comparison to DLS data
\cite{porter97a}. The systems $d+$Ca, $\alpha+$Ca and
Ca+Ca as well as C+C
have been examined. See
figs. \ref{dlspp} and \ref{dlspd} for additional
information.
 \item[ FIG. \ref{ca1ca} ] Total dilepton spectra from Ca+Ca
collisions compared to two generations of data. To compare
to the earlier measurements \cite{roche89a} the
old DLS filter V1.5 (dashed line) has been
applied. The solid line corresponds
to the calculation of fig.~\ref{dlsaa}.
 \item[ FIG. \ref{camed} ] Total dilepton spectra from Ca+Ca
collisions with an increased $\omega$ width. Also shown are
effects due to density dependent modifications of the
vector meson masses.
  \end{description}
\newpage
{\Large Fig. \ref{ppc}}
\newpage
{\Large Fig. \ref{dalspectra}}
{\Large Fig. \ref{bremsdm}}
\newpage
{\Large Fig. \ref{dlspp}}
\newpage
{\Large Fig. \ref{dlspd}}
\newpage
{\Large Fig. \ref{pd2pp}}
\newpage
{\Large Fig. \ref{dlsaa}}
\newpage
{\Large Fig. \ref{ca1ca}}
{\Large Fig. \ref{camed}}

\end{document}